\documentclass[12pt]{article}
\usepackage{amsmath}
\usepackage{amssymb}
\usepackage[spanish]{babel}
\usepackage[latin1]{inputenc}

\begin{document}
\begin{center}
{\Large {\bf  $GL(3,R)$ gauge theory of gravity coupled\\ with an
electromagnetic field }} \vskip 10pt {\bf Rolando Gaitan}${}^{a,
}${\footnote {e-mail: rgaitan@fisica.ciens.ucv.ve}}
and {\bf Frank Vera}\\
${}^a${\it  Departamento de F\'\i
sica, Facultad Experimental de Ciencias y Tecnolog\'{\i}a, \\
Universidad de Carabobo, A.P. 129
Valencia 2001, Edo. Carabobo, Venezuela.}\\
\end{center}
\vskip .1truein

\begin{center}
\begin{minipage}{5in}\footnotesize\baselineskip=10pt
    \parindent=0pt

Consistency of $GL(3,R)$ gauge theory of gravity coupled with an
external electromagnetic field, is studied. It is shown that
possible restrictions on Maxwell field can be avoided through
introduction of auxiliary fields.

\end{minipage}
\end{center}





{\bf 1) Introduction }

Among several gauge formulations for gravity, we consider a
Yang-Mills type with the Lie group $GL(3,R)$ as a gauge
group$^{1,2}$, where a covariant coupling scheme is known with
material fields in a space-time with non null torsion
provided$^{2}$. Here, we focus our attention on consistency
between Einstein-Hilbert Theory (EHT) and a $GL(3,R)$ gauge theory
of gravity coupled with an external electromagnetic field, in a
$2+1$ dimensional space-time, not necessarily with null torsion.
There, the standard definition of Maxwell field must depend on
vector potential and torsion, so the source of gravitation, given
by the electromagnetic energy-momentum tensor depends on
equivalence class of connection. This fact conduces to an
extension of model discussed in reference$^{2}$ and constitutes a
particular type of a lagrangian formulation with four order
self-interaction terms on connection.

This note is organized as follows. In section 2 we review a gauge
formulation of gravity based on a frame bundle. Next, the coupling
between gravity, Maxwell and auxiliary fields is performed,
confirming consistency with EHT at the torsionless limit. We
conclude with some remarks.

\vskip .1truein

{\bf 2) $GL(3,R)$ gauge formulation for gravity }

Let be $\mathcal{M}$ a $GL(3,R)$ frame bundle on a 2+1 space-time
as a base space with a metric $g_{\mu \nu }$ and coordinates
$x^\mu $ provided. The connection 1-form shall be considered as an
independent object from metric and it is introduced through
${(\mathbb{A}_\lambda)^\mu}_\nu \equiv {\Gamma^\mu}_{\lambda\nu}$,
where ${\Gamma^\mu}_{\lambda\nu}$  is the affine connection which
allows the construction of the covariant derivative, $\nabla_\mu$,
the torsion tensor,
${T^\mu}_{\lambda\nu}={(\mathbb{A}_\lambda)^\mu}_\nu
-{(\mathbb{A}_\nu)^\mu}_\lambda $ and the curvature,
${R^\sigma}_{\alpha \mu \nu }\equiv {(\mathbb{F}_{\mu \nu})^\sigma
}_\alpha={\big (\partial _\nu \mathbb{A}_\mu -
\partial _\mu \mathbb{A}_\nu+
[\mathbb{A}_\mu\,,\,\mathbb{A}_\nu]\big)^\sigma}_\alpha$. So, the
$GL(3,R)$ gauge invariant action for the free model at the
torsionless limit, is
\begin{equation}
S_o = \kappa \,\int_{\mathcal{M}} d^3 x \sqrt{-g} \,\, \big(-\frac
{1}4 tr\,
\mathbb{F}^{\alpha\beta}\mathbb{F}_{\alpha\beta}+C_{\alpha\beta}\,
\varepsilon^{\beta\lambda\sigma}{(\mathbb{A}_\lambda)^\alpha}_\sigma\big)
\,\, , \label{2}
\end{equation}
where $\kappa$ has a length dimension and $C_{\alpha\beta}$ are
the lagrange multipliers related to torsion constraint. This model
could be cosmologically extended with the introduction of a
lagrangian term ''$\lambda ^2$'' (see reference$^{2}$), where
$\lambda$ is the cosmological constant. Variations on connection
and metric provide the field equations, $\nabla_\alpha R_{\lambda
\sigma } - \nabla_\sigma R_{\lambda \alpha } =0$ and
$2R_{\sigma\mu}{R^\sigma}_\nu
-2RR_{\mu\nu}-g_{\mu\nu}\,R_{\sigma\lambda}R^{\sigma\lambda}+\frac{3}{4}\,g_{\mu\nu}\,R^2=0$,
which conduce to a flat solution ($\mathbb{F}_{\alpha\beta}=0$),
in consistency with free Einstein theory in 2+1 dimensions.
\vskip .1truein

{\bf 3) Coupling with Maxwell and auxiliary fields }

It is well known that electromagnetic field couples with torsion
because, if $a_\mu$ is the potential trivector, the Maxwell tensor
is
\begin{equation}
f_{\mu\nu}=f^{(o)}_{\mu\nu}+a_\lambda
\varepsilon^{\rho\alpha\beta}
\varepsilon_{\rho\nu\mu}{(\mathbb{A}_\alpha)^\lambda}_\beta\,\, ,
\label{6}
\end{equation}
where $f^{(o)}_{\mu\nu}=\partial_\mu a_\nu -\partial_\nu a_\mu$
and $\varepsilon^{\rho\alpha\beta}$ is the Levi-Civita (pseudo)
tensor. Using (\ref{6}), the Maxwell lagrangian density
($\mathcal{L}_M$) and symmetric momentum-energy tensor
($T^{M}_{\mu\nu}$) can be defined and they are given by
\begin{equation}
\mathcal{L}_M=-\frac{1}{4}\,\psi \,\, , \label{6a}
\end{equation}
\begin{equation}
T^{M}_{\mu\nu}=\psi_{\mu\nu}-\frac{g_{\mu\nu}}{4}\,\psi \,\, ,
\label{6b}
\end{equation}
where notation means $\psi_{\mu\nu}\equiv
f_{\mu\sigma}{f_\nu}^\sigma$ and $\psi\equiv
f_{\mu\sigma}f^{\mu\sigma}$.

Following reference$^{2}$ and using (\ref{2}) we introduce the non
minimal coupling between gravity, electromagnetism and auxiliary
connection ($\mathbb{W}_\alpha$), at the torsionless limit
\begin{eqnarray}
S=S_o +\kappa \int_{\mathcal{M}} d^3 x \sqrt{-g}\big(\ell(g,f) -
8\pi G\,tr\,\mathbb{M}^{\alpha}(g,f)\,^*\mathbb{F}_{\alpha}
+tr\,\mathbb{J}^\alpha(\mathbb{A}_\alpha-\mathbb{W}_\alpha)
\nonumber \\+\,tr\,\mathbb{H}^{\alpha
\beta}(\mathbb{A}_\alpha-\mathbb{W}_\alpha
)(\mathbb{A}_\beta-\mathbb{W}_\beta)\big)
 \,\, \,,\label{7}
\end{eqnarray}
where $\ell(g,f)$ is the modified Maxwell lagrangian density with
second order correction given by parameters $b_1$ and $b_2$
\begin{equation}
\ell(g,f)\equiv  \mathcal{L}_M +b_1 (T^{M})^2 + b_2 T^{M}_{\mu
\nu}{T^{M}}^{\mu \nu}
 \, \,. \label{8}
\end{equation}

The components of the coupling tensor, $\mathbb{M}^\alpha$ have
linear dependence on $T^{M}_{\mu\nu}$, and they are
\begin{eqnarray}
{(\mathbb{M}^\alpha )^\mu }_\nu = \big[c_1
{\varepsilon^{\alpha\mu}}_\nu g^{\sigma \rho}
 + c_2{\varepsilon^{\alpha\rho}}_\nu g^{\sigma \mu}+ c_3\varepsilon^{\alpha\mu\rho}
{\delta^\sigma }_\nu\big]\,T^{M}_{\sigma \rho} + a
{\varepsilon^{\alpha\mu}}_\nu\, \,, \label{9}
\end{eqnarray}
where $c_1$, $c_2$, $c_3$ and $a$ are free parameters, and
$^*\mathbb{F}_{\alpha}$ in (\ref{7}) is the Poincar\'e dual of
curvature (i.e., $^*\mathbb{F}_{\alpha}\equiv
\frac{1}{2}\varepsilon_{\alpha\mu\nu}\mathbb{F}^{\mu\nu}$).

The coupling tensors for auxiliary fields are
\begin{equation}
(\mathbb{J}_{\beta})_{\mu \nu }\equiv (d_1 + d_2 \psi){\varepsilon
}_{\beta \mu \nu} \, \,, \label{10}
\end{equation}
\begin{equation}
(\mathbb{H}^{\alpha \beta})^{\mu \nu} \equiv a_1
\,g^{\alpha\beta}g^{\mu\nu}+a_2 \,g^{\alpha\mu}g^{\beta\nu}+ a_3
\,g^{\alpha\nu}g^{\beta\mu}
 \, \,, \label{11}
\end{equation}
where $d_1$, $d_2$, $a_1$, $a_2$ and $a_3$ are free parameters and
$\psi\equiv f_{\mu\sigma}f^{\mu\sigma}$.

We underline some aspects of action (\ref{7}). The lagrangian
density $\ell(g,f)$ encloses Maxwell density and second order
contributions on momentum-energy tensor, this means second order
in $\psi_{\mu\nu}$. On the other hand, the lagrangian term ''$8\pi
G\,tr\,\mathbb{M}^{\alpha}\,^*\mathbb{F}_{\alpha}$'' is
covariantly defined and provide a minimal coupling contribution
plus non minimal ''Proca'' density. This fact suggests the shape
of coupling with auxiliary fields $\mathbb{W}_\alpha$.

The field equation for $\mathbb{W}_\alpha$ is $\mathbb{J}^\beta +
\mathbb{H}^{\alpha
\beta}(\mathbb{A}_\alpha-\mathbb{W}_\alpha)+(\mathbb{A}_\alpha-\mathbb{W}_\alpha)
\mathbb{H}^{\beta\alpha}=0$, so an ansatz for $\mathbb{W}_\alpha$
is chosen, $(\mathbb{A}_{\alpha}-\mathbb{W}_{\alpha})_{\mu \nu}=
(\theta _1 + \theta _2 \psi ) \,{\varepsilon }_{\alpha\mu\nu}$
with $d_n = 2a_{21}\theta _n$ for $n=1,2$ and $a_{21} \equiv
a_2-a_1$. Then, the field equation for $GL(3,R)$ connection can be
written like
\begin{eqnarray}
\nabla_{\nu}\big(R^{\alpha \mu}-8\pi G c_1 g^{\alpha \mu}{T^{M}}
-8\pi G c_2{T^{M}}^{\alpha \mu} \big)-
\nabla^{\mu}\big({R^{\alpha}}_{\nu}-8\pi G c_1
{{\delta}^{\alpha}}_{\nu}{T^{M}} -8\pi G
c_3{{T^{M}}^{\alpha}}_{\nu}\big)\nonumber \\+\,8\pi G
\nabla_\beta\big(c_2 {{\delta}^{\alpha}}_{\nu}  {T^{M}}^{\mu
\beta} -c_3g^{\alpha \mu}
{{T^{M}}^{\beta}}_{\nu}\big)+\kappa^{-3}\varepsilon^{\lambda\alpha\mu}
\widetilde{C}_{\lambda\nu}=0 \, \,, \label{12}
\end{eqnarray}
where the lagrange multipliers have been rewritten as
$\widetilde{C}_{\lambda\nu}\equiv
C_{\lambda\nu}+(1-2(b_1-b_2)T^{M}-88\pi GR)a_\nu \,^*f_\lambda
-8(b_2{{T^M}^\rho}_\mu +8\pi G
{R^\rho}_\mu)\varepsilon_{\lambda\sigma\rho}a_\nu f^{\mu\sigma}$.

Demanding consistency between (\ref{12}) and field equation of EHT
(i.e., $R_{\alpha \beta }-\frac{g_{\alpha \beta }}2R=-8\pi G
{T^{M}}_{\alpha \beta }$), some parameters are fixed,
$c_1=-c_2=-c_3=1$ and ${T^{M}}_{\alpha \beta }$ must be a
covariantly conserved tensor.

Variations on metric of (\ref{7}) provide an equation whose
evaluation on EHT conduce to
\begin{eqnarray}
\big(-8b_1-24b_2+24(8\pi G)^2 -768a_{21}\theta_1^2\big)\psi \psi
^{\rho\sigma}\nonumber \\
+\big(b_1+3b_2-17(8\pi G)^2 +32a_{21}\theta_2^2\big){\psi}^2g^{\rho\sigma}   \nonumber \\
+\big(16-32(8\pi G)^2a
-768a_{21}\theta_1\theta_2\big){\psi}^{\rho\sigma}\nonumber \\
+\big(-4+8(8\pi G)^2a +192a_{21}\theta_1\theta_2\big)\psi
g^{\rho\sigma} \nonumber \\
+\,96a_{21}\theta_1^2g^{\rho\sigma} =0 \, \,, \label{13}
\end{eqnarray}
and for all $\psi_{\mu\nu}\equiv f_{\mu\sigma}{f_\nu}^\sigma$, the
remaining free parameters can be fixed
\begin{equation}
b_1+3b_2=3(8\pi G)^2
 \, \,, \label{14a}
\end{equation}
\begin{equation}
a_{21}\theta_2^2=14(8\pi G)^2
 \, \,, \label{14b}
\end{equation}
\begin{equation}
\theta_1=0
 \, \,, \label{14c}
\end{equation}
\begin{equation}
2(8\pi G)^2a=1
 \, \,, \label{14d}
\end{equation}
so equation (\ref{13}) is satisfied identically.

\vskip .1truein

{\bf 4) Concluding remark}

The non minimal coupling scheme for the Yang-Mills-like gauge
formulation of gravity provides the necessary terms which
guarantee the fixing of free parameters in order to obtain
consistency between $GL(3,R)$ connection and EHT. This fact says
that a minimal coupling is not sufficient. However, metric field
equation demands the presence of auxiliary connection, on the
contrary, quadratic polynomial restrictions over $\psi_{\mu\nu}$
would appear. It is possible to choose parameters $b_1$ and $b_2$
in the way that $\ell(g,f)\rightarrow \mathcal{L}_M$, when gravity
decouples ($G\rightarrow 0$), so one can recover the free Maxwell
theory in a flat space-time.

\vskip .1truein

{{\bf  Acknowledgment }}

This work is partially supported by project CDCH-UC 1102-06. We
thank Prof. Ilse Rodriguez for technical support.


\vskip .1truein

{\bf  References }

[1] F. Mansouri, L. N. Chang, {\it Phys. Rev. }{\bf D13}, No. 12
(1976) 3192.

[2] R. Gaitan, {\it Mod. Phys. Lett. }{\bf A}, Vol. 18, No. 25
(2003) 1753.





\end{document}